\documentstyle[12pt]{article}
\begin{document}

\begin{titlepage}
\date{January 20, 2000}
\title{Stochastic Processes and Thermodynamics \\ on Curved Spaces}
\author{ Sergiu I.\ Vacaru \thanks{\copyright\ Sergiu I. Vacaru,\ e-mail:
vacaru@lises.asm.md}  \\[10pt]
  \small Institute of Applied Physics,  Academy of Sciences,\\
  \small Academy str. 5,\  Chi\c sin\v au MD2028, \\
  \small Republic of Moldova     }
\maketitle
\vskip0.2cm
Elecronic preprint: {\bf gr--qc/0001aaa}
\vskip 2 cm
\thispagestyle{empty}
\begin{abstract}
\baselineskip18pt
 Our approach views the thermodynamics and kinetics
  in general relativity and extended gravitational theories
 (with generic local anisotropy) from the perspective of the
  theory of stochastic  differential equations  on curved spaces.
   Nonequilibrium and irreversible processes in
  black hole thermodynamics are considered. The paper summarizes
 the author's contribution to Journees Relativistes 99 (12--17 September
 1999), Weimar, Germany.
 \end{abstract}
\end{titlepage}
\section{Diffusion, Kinetics and Thermodynamics  \newline
 in Locally Isotropic and Anisotropic \newline spacetimes }
We generalized  \cite{v5} the  stochastic calculus  on Riemannian
 manifolds  for  an\-isot\-rop\-ic processes anf for  fiber bundles
 provided with nonlinear connection structure \cite{v1}.
 Lifts in the total space of linear  frame bundles were used in order
 to consider Browinian motions, Wiener  processes and Langevin
 equations in a covariant fashion. The concept  of thermodynamic
 Markovicity and Chapman--Kolmogorov  equations were analyzed in
 connection to the possibility of obtaining
 information about pair--correlation functions on curved spaces.

 Fokker--Plank type  covariant equations were derived for both locally
 isot\-rop\-ic and anisotropic gravitational and matter field interactions.
 Stability of equilibrium and nonequilibrium states,
 evolution criteria, fluctuations and
 dissipation are examinded from the view point of a general
 stochastic formalism on curved spaces.

 The interrelation between  classical statistical mechanics,
 thermodynamics and kinetic theory  (the Bogolyubov --- Born and
  Green --- Kirkwood --- Yvon herachy,  and derivation of Vlasov and
 Boltzmann equations \cite{vlas}) was studied  on Riemannian manifolds
  and vector bundles.

The covariant diffusion and hydrodynamical approximations \cite{sy},
 the kinematics
 of relativistic processes, transfering and production of entropy,
  dynamical equations and thermodynamic relations were
 consequently defined. Relativistic formulations \cite{groot}
 and anisotrop\-ic  generalizations were considered for extended
 irreversible thermodynamics.

\section{Thermodynamics of Black Holes with Local Anisotropy}

The formalism outlined in the previous section was applied to
 cosmological models and black holes with  local spacetime anisotropy
 \cite{v3,v4}.

  We analyzed the conditions when the Einstein equations with cosmological
 constant and matter (in  general relativity and low dimenensional and
 extended variants of gravity)  describe generic locally anisotropic
 (la) spacetimes. Following De Witt  approach we set up a method for
 deriving energy momentum tensors for locally anisotropic matter.

We speculated on black la--hole solutions induced by locally anisotropic
 splittings  from tetradic, spinor and gauge and ge\-ne\-ra\-liz\-ed
 Kaluza--Klein--Fins\-ler models  of gravity  \cite{v1,vg}.
  Possible extensions  of la--metrics
 \cite{v2,v3} to string and brane models were considered.

The thermodynamics of (2+1) dimensional
black la--holes was discussed in connection with a
 possible statistical mechanics background based on locally anisotropic
 variants of Chern--Simons theories \cite{v4}. We proposed a variant of
 irreversible thermodynamics for black la--holes. There were also
 considered  constructions and calculus of thermodynamic
 parameters of black la--holes, in the framework of
 approaches to thermodynamic geometry \cite{rup}
 for nearly equilibrium states,
 and the effects of local nonequilibrium and questons of stability
 were analyzed by using  thermodynamic metrics and curvatures.
\vskip0.3cm
{\bf Acknowledgements:}\  The author thanks
 the Organizers and Deutsche Forschungsgemeinschaft for kind
 hospitality and support of his participation at Journees
 Relativistes 99.

\end{document}